\documentclass{ws-ijmpd}

\begin{document}

\markboth{Z. B. Zhang et al.} {Probing the emission regions for
distinct gamma-ray bursts}

%
\catchline{}{}{}{}{}
%

\title{Inferring the emission regions for different kinds of gamma-ray bursts}

\author{Zhibin Zhang$^{1, 3, \dag}$, G. Z. Xie$^{1}$, J. G. Deng$^{2}$ and B. T. Wei$^{1, 3}$}

\address{$^1$ National Astronomical Observatories/Yunnan Observatory,
Chinese Academy of Sciences\\ P. O. Box 110, Kunming, Yunnan,
650011,
P. R. China\\
$^2$ Physics Science and Technology Institute, Guangxi University\\
Nanning, Guangxi 530004, P. R. China\\
$^3$ The Graduate School of the Chinese Academy of Sciences\\
$\dag$ zbzhang@ynao.ac.cn}

%

\maketitle

\begin{history}
\received{Day Month Year}
\revised{Day Month Year}
\comby{Managing Editor}
\end{history}

\begin{abstract}
Using a theoretical model describing pulse shapes, we have clarified
the relations between the observed pulses and their corresponding
timescales, such as angular spreading time, dynamic time as well as
cooling time. We find that the angular spreading timescale caused by
curvature effect of fireball surface only contributes to the falling
part of the observed pulses, while the dynamic one in the co-moving
frame of the shell merely contributes to the rising portion of them
provided the radiative time is negligible. In addition, the observed
pulses resulted from the pure radiative cooling time of relativistic
electrons exhibit the property of fast rise and slow decay (a
quasi-FRED profile) together with smooth peaks. Besides, we
interpret the phenomena of wider pules tending to be more asymmetric
to be a consequence of the difference in emission regions.
Meanwhile, we find the intrinsic emission time is decided by the
ratios of lorentz factors and radii of the shells between short and
long bursts. Based on the analysis of asymmetry, our results suggest
that the long GRB pulses may occur in the regions with larger
radius, while the short bursts could locate at the smaller distance
from central engine.
\end{abstract}


\section{Introduction}

Owing to possible overlapping between neighboring pulses, time
profiles of gamma-ray bursts (GRBs) usually exhibit very diverse
configurations$^{18}$. Pulses as the fundamental elements of bursts
provide us a clue which will help us catch their quiddities.
According to the standard fireball shock model$^{19, 33, 15}$, the
pulses are generally proposed to either originate from the internal
shocks produced by collisions between different shells in the
fireball$^{24, 37}$ or come from the action of the external shock on
the circum-burst medium$^{14, 26, 4, 6}$. However, problems are
still in existence not only for their radiation mechanisms but also
at the aspect of the differences between short and long pulses in
their profiles.

It had been pointed out that GRB light curves especially the pulses
in them could reproduce the temporal activity of its inner
engine$^{10, 2, 17}$. Dermer \& Mitman$^{5}$ had found that
investigations on pulses were preferred for learning whether GRB
sources require engines to be long-lasting or short-lived. The
long-lasting central engine should produce more relativistic
outflows during this phase which naturally lead to long-lasting
emissions in the co-moving frame of the shells. The long time
together with the delay time caused by the curvature effect of
photosphere can inevitably results in longer duration in the
observer frame$^{39,}$\footnote{hereafter paper I}. Qin and
Lu$^{23}$ thought that for most of observed pulses their
corresponding co-moving pulses might contain a long decay time
relative to the time-scale of the curvature effect. Moreover,
studies with BATSE data for GRB pulses also indicate most sources do
not display the curvature effects$^{11, 1}$, which suggests that the
physics of such pulse formation is dominated or determined by other
effects, perhaps the hydrodynamic or radiative cooling process or
both of them. It happens that the relations between these timescales
and temporal structure of GRBs had been researched by some
authors$^{27, 29}$.

For the long and smooth single-peak GRBs, the curvature
effects$^{9}$ or external shock$^{4}$ could be the leading action.
The result that many short bursts are highly variable demonstrates
short GRBs can't be produced by external shocks$^{16}$. So one of
the aims in this work is to conjecture the approximate regions where
bursts could occur. With this purpose, the structures of observed
pulses connecting with different timescales are taken into account
in very detail. These results suggest that short and long bursts
might take place in different spatial regions from the sources.

\section{Theoretical model}

In this section we first introduce an analytic model based on the
assumption of isotropic radiation in the frame of
fireball$^{22,}$\footnote{hereafter paper II}. The model has offered
us a fundamental expression of pulses concerning observed counts
flux (see paper II, eq. [21])
\begin{equation}
C(\tau)=C_0\frac{\int\limits_{\widetilde{\tau}_{\theta,min}}^{\widetilde{\tau}_{\theta,max}}\widetilde{I}(\tau_\theta)(1+\beta\tau_\theta)^2(1-\tau+\tau_\theta)d\tau_\theta
}{\Gamma^3(1-\beta)^2(1+\frac{\beta}{1-\beta}\tau)^2}
\int\limits_{\nu_1}^{\nu_2}\frac{g_{0,
\nu}(\nu_{0,\theta})}{\nu}d\nu
\end{equation}
with
\begin{equation}
\tau=\frac{T-t_{c}-D/c+R_{c}/c}{R_{c}/c}
\end{equation}
and
\begin{equation}
\tau_{\theta}=\frac{t_{\theta}-t_{c}}{R_{c}/c}
\end{equation}
where $R_{c}$ and $\Gamma$ denote the corresponding initial radius
of the fireball at time $t_c$ and the lorentz factor of ejecta when
photons start emitting from the shocked shells, and $T$ and
$t_{\theta}$ stand for the observation time by observer and the
emission time from the photosphere respectively. $D$ is the
luminosity distance from observer to the source. $\tau$ and
$\tau_{\theta}$ are two dimensionless quantities and related by the
following expression
\begin{equation}
\tau_{\theta}=\frac{\tau-(1-\mu)}{1-\beta\mu}
\end{equation}
where $\mu=cos\theta$, and more detailed definitions of the relevant
variables can be referred to paper I and II. Followed our recent
work (paper I) the co-moving pulse form is assumed to be
$$
\widetilde{I}(\tau_\theta)\equiv I(\tau_\theta)=I_0\left\{
\begin{array}{cc}
exp[(\tau_{\theta}-\tau_{\theta,0})/\sigma_{r}] \ & \ \ (\tau_{\theta,min}\leq\tau_{\theta}\leq\tau_{\theta,0})\\
exp[-(\tau_{\theta}-\tau_{\theta,0})/\sigma_{d}] \ & \ \ (\tau_{\theta,0}<\tau_{\theta}\leq\tau_{\theta,max})\\
\end{array}
\right.
$$
and let $\sigma_{r}/\sigma_{d}=\xi$.

From eq.(1), the one-to-one relation between $C(\tau)$ and $\tau$
can be constructed on condition that the values of the lorentz
factor $\Gamma$ and the co-moving width
$\Delta\tau_{\theta}=\tau_{\theta, max}-\tau_{\theta, min}$ have
been assigned in advance.

\section{Time scales}

It has been known that the structure of GRB pulse is usually thought
to be determined by three time scales. The first is angular
spreading timescale, $T_{ang}\approx R_{c}/(2\Gamma^{2}c)$, which is
caused by the interval between the arrival times of the photons
emitted from different region of the shell$^{28}$. The second is
dynamic timescale or shock shell-crossing time, $T_{dyn}\approx
t_{dyn}^{'}/2\Gamma$ with $t_{dyn}^{'}=\Delta^{'}/v_{sh}^{'}$ in the
rest frame of shell$^{25}$, where $\Delta^{'}$ is the thickness of
shell and $v_{sh}^{'}$ is the velocity of shock relative to the
pre-shocked flow. The last one is radiative generally synchrotron
cooling time, $T_{syn}\simeq t_{\gamma}^{'}/\Gamma$, where
$t_{\gamma}^{'}$ is the radiative timescale in co-moving frame of
the shell$^{32}$. As the duration the asymmetry of pulses (For
convenience, we define the asymmetry in previous manner as the ratio
of the rise fraction ($t_r$) of full width at half maximum ($FWHM$)
of pulse to the decay fraction ($t_d$).) is likewise thought to be
decided by the above three timescales.

Now, Let us consider an uniform and spherical shell moving outwards
with a radial velocity of $v=c\beta$, which emits photons
consecutively at different times as the shock propagates into it.
The relation between observation time $T$ and the emitting time $t'$
in co-moving frame of the shell has been gotten$^{31}$ and can be
written as
\begin{equation}
T\equiv(1-\beta\mu)t'+(1-\mu)R_{c}/c=T_{int}+T_{ang}
\end{equation}
where the first part of the right-hand side of the equation
resulting from the time $t'$ is defined as intrinsic timescale, and
the second part is purely caused by the curvature effects. Assuming
the distance of observer from the central engine is $D$, from
equations (2), (4) and (5), one can verify the meaning of $T$ in
eqs. (2) and (5) is identical if only I let the initial time at
radius $R_c$ be constrained with
\begin{math}
t_{c}=(R_{c}-D)/c.
\end{math}
In this situation, I can get $t'=\Delta\tau_{\theta}R_{c}/c$, where
$\Delta\tau_{\theta}=\sigma_{r}+\sigma_{d}$ denotes the width of
co-moving pulse (see paper I). Then I find that
\begin{equation}
T_{int}=(1-\beta\mu)t'=(1-\beta\mu)\Delta\tau_{\theta}R_{c}/c
\end{equation}
actually involves the contributions of timescales arising from
co-moving frame to the pulse duration. In terms of the current
fireball shock model, the intrinsic time $T_{int}$ (or
$\Delta\tau_{\theta}$) is often interpreted as a consequence of
combination of $T_{dyn}$ and $T_{syn}$.

\section{Resulting pulses}

We examine the dependence of the structure of pulses on the distinct
timescales when parts of them are taken no account of due to some
special physical reasons. Suppose the rising part of the co-moving
pulse corresponds to the shell crossing time and its decay portion
relates to the cooling time, we thus have the opportunity to
distinguish their individual effects on the observed temporal
profiles in the following extreme instances. In the internal shock
model, pulses are caused by the dissipation of a fraction of kinetic
energy of ejecta with lorentz factors that could exceed 100$^{8,
19}$. In the following, we take $\Gamma=100$ as the representative
value for the most sources.

\subsection{Pulses dominated by $T_{ang}$}

If the shell thickness $\Delta'$ becomes thin enough due to the
collision between the shell and the circum-burst medium at larger
distance from the central engine, and/or the velocity of the forward
shock is relativistic$^{26}$, the intrinsic time $T_{int}$ would be
expected to be very small relative to $T_{ang}$. In this case, the
effective timescales in eq. (5) are then reduced to be
\begin{equation}
T\approx T_{ang}=(1-\mu)R_{c}/c\approx R_{c}/2\Gamma^{2}c
\end{equation}
in the case of $\Gamma\gg 1$.
\begin{figure}
\centering
 \includegraphics[width=5in,angle=0]{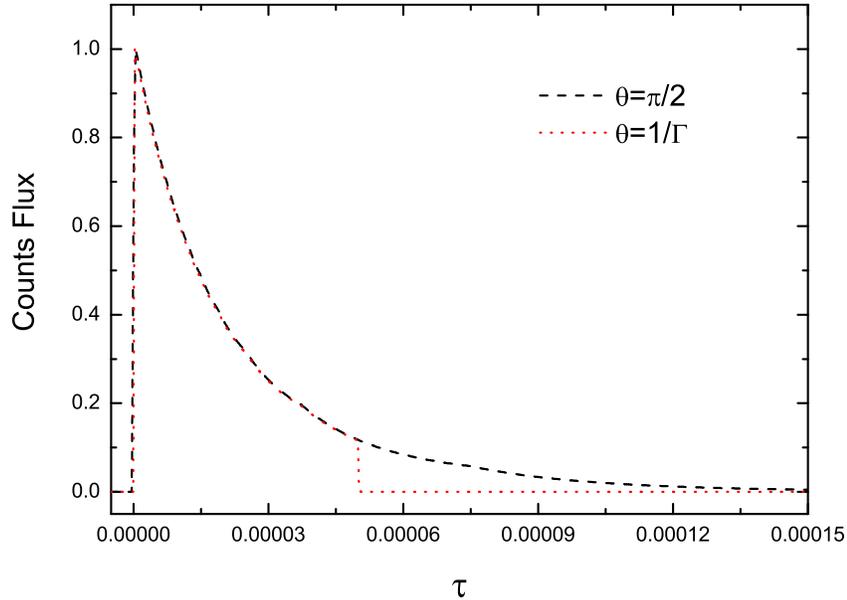}
    \caption{Plot of counts flux vs. $\tau$ for the normalized curves from angular spreading time, where we take $\Gamma=100$,
     $\Delta \tau_{\theta}=0.001$ and $R_c=10^{15} cm$. The dashed and dotted lines represent
     different emitting areas on the fireball surface and have been symbolized in the plot.}
  \label{fig1}
 \end{figure}

The resulting pulses have been displayed in figure 1. Here we can
optionally appoint the value of $\xi$, say, for special $\xi=1$. In
fact, our previous work (see, paper II) has concluded that the
expected pulses are independent of their co-moving pulse shapes when
their local width is narrow enough. We find from figure 1 that the
pulses resulted from the angle spreading timescale behave the
standard decay form in the shape\footnote{Paper II had proved
     the resulting pulse shapes tend to follow the so-called standard decay form and be independent of
     co-moving pulse forms when the co-moving width ($\Delta\tau_{\theta}$) is short
     enough.}, although they come from distinct surface.
Furthermore, the discrepancy in the two durations demonstrates the
contributions of photons from photosphere to light curves decrease
with the reduction of the emitting area.

\subsection{Pulses arising from $T_{dyn}$}

Considering an ejecta moving towards us in the line of sight or
beaming effect due to large lorentz factors in a small distance, the
curvature effect on the observation might become very tiny. In the
case of neglecting relativistic curvature effect (i.e. $\mu=1$), the
dominant component of the duration $T$ in eq. (5) would remain to be
\begin{equation}
T\approx T_{int}=T_{dyn}=(1-\beta\mu)t'\approx
\Delta\tau_{\theta}R_{c}/2\Gamma^{2}c
\end{equation}
where the $\Delta\tau_{\theta}$, $R_c$ and $\Gamma$ are taken some
typical values as 10, $10^{13} cm$ and 100 respectively. To contrast
the effects of diverse ratios between $\sigma_r$ and $\sigma_d$ on
observed pulse shapes, we assign $\xi=3.0, 5.7$ and $9.0$
correspondingly. Here the angle is taken as $\theta=1/(10^5\Gamma)$
so that it is ensured to be small enough. The current cases
represent fast cooling physical process in co-moving frame of the
shell.

In the following, for each pulses, the magnitude of count fluxes is
normalized to a unit, and the relative time $\tau$ was re-scaled so
that their right end of $FWHM$ is located at the same point. These
normalized and re-scaled curves have been displayed in figure 2,
from which we can find that the co-moving pulses with dominant
dynamic timescales would lead to spiky analytic pulses. At the same
time, the asymmetry of analytic pulses is proportional to the
co-moving pulses' ratio. In other words, with the increasing of
dynamic timescale the observed pulses would become more asymmetric
and show like exponential rise and fast decay (ERFD). Under extreme
physical conditions, we shall draw the conclusion that the unmixed
dynamic times should result in highly spiky pulses without decay
portion.
\begin{figure}
\centering
 \includegraphics[width=5in,angle=0]{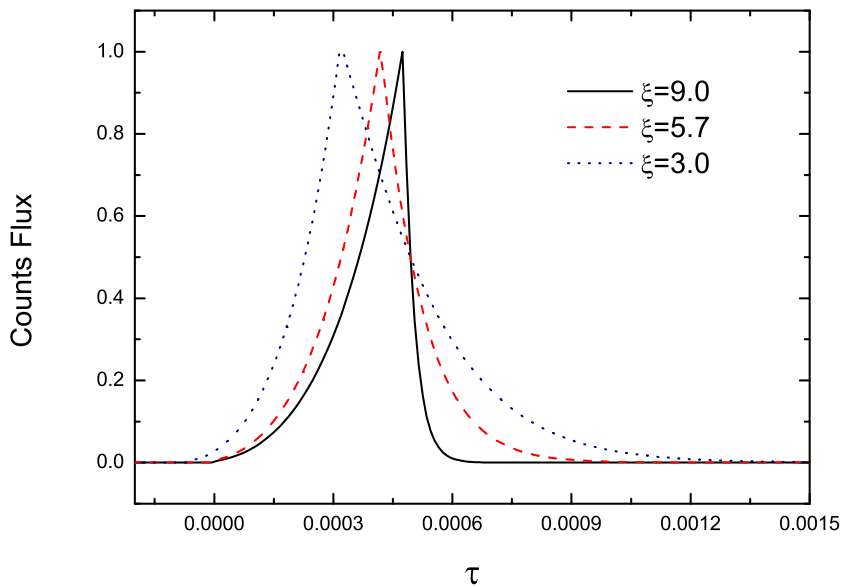}
    \caption{Plot of counts flux vs. $\tau$ for the normalized and re-scaled curves from dynamic timescale, where we take $\Gamma=100$,
     $\Delta \tau_{\theta}=10$, $R_c=10^{13} cm$ and $\theta=1/(10^5\Gamma)$. Different symbols have been shown in this plot.}
  \label{fig2}
 \end{figure}

\subsection{Pulses originated from $T_{syn}$}

As pointed by previous authors$^{32, 30}$, the cooling time-scale is
probably much larger than the above two time-scales, for instance,
the accelerated particles radiate quite slowly, especially for very
large radii due to the low densities of the shell$^{17}$. It is thus
necessary to reveal how the observed curves are influenced by the
radiation of electrons. The cooling timescale can be expressed as
follows
\begin{equation}
T\approx T_{int}=T_{syn}=(1-\beta\mu)t'\approx
\Delta\tau_{\theta}R_{c}/2\Gamma^{2}c
\end{equation}
where the co-moving pulses must be mostly formed by its decay
potion.

Resembling figure 2, the normalized and re-scaled curves from eq.
(1) for different co-moving pulses' ratios, say $\xi=1.0, 0.3, 0.1$,
are similarly plotted in figure 3. On the occasions of $\xi \ll 1$,
the analytic pulses are almost governed by radiative cooling time,
namely considerably slow cooling process comparable to dynamical
timescale of shocks crossing the shocked flows. From figure 3 we
find the shapes of all these resulting curves in this case follow a
form of fast rise and slow decay (quasi-FRED), and a profile of
smooth instead of spiky peak. Additionally, with the increasing of
cooling timescale the observed pulses would become more symmetric
and smooth. Taking into account of an ultimate situation, $\xi =0$,
we calculate the asymmetry of the resulting pulse and find it has an
upper limit, here 0.37, when the parameters are assigned to be
certain values such as $\Gamma=100$, $\Delta \tau_{\theta}=10$ and
$R_c=10^{13} cm$. The theorem certainly holds for other curves from
any of sets of parameters, provided the curves are completely
contributed by the cooling time.
 \begin{figure}
\centering
 \includegraphics[width=5in,angle=0]{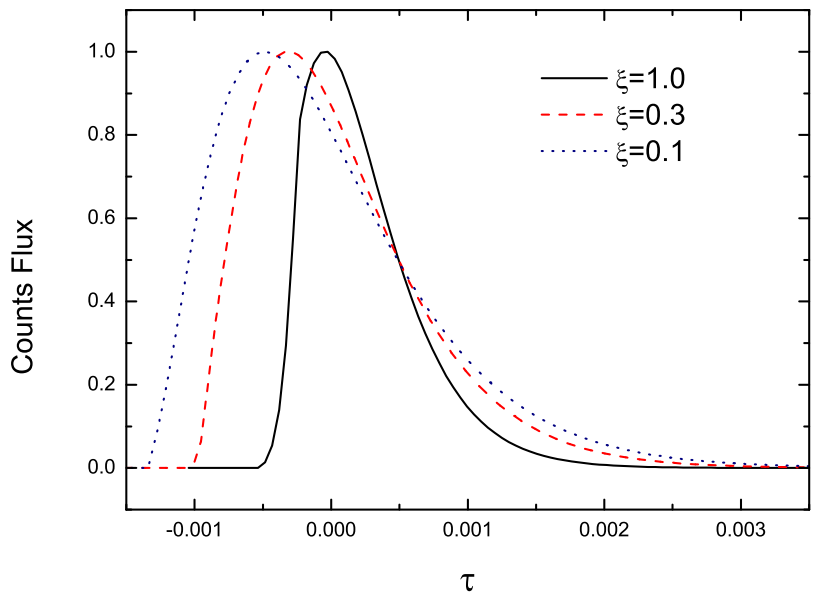}
    \caption{Plot of counts flux vs. $\tau$ for the normalized and re-scaled curves from radiative cooling time, where we take $\Gamma=100$,
     $\Delta \tau_{\theta}=10$, $R_c=10^{13} cm$ and $\theta=1/(10^5\Gamma)$. Different symbols have been shown in this plot.}
  \label{fig3}
 \end{figure}

\section{Comparison between $T_{ang}$ and $T_{dyn}$}

In fact, for most reasonable parameters the cooling time is much
shorter than other physical timescales$^{27, 10, 19, 34}$,
especially in the scenario of internal shocks. In this case,
$T_{ang}$ and $T_{dyn}$ could be the key factors acting on the
properties of observed pulses. To discern which one is more
significant than the other, I contrast the two timescales from eqs.
(7) and (8) in figure 4. The reason for taking $\theta\sim1/\Gamma$
in $T_{ang}$ is that the outflows crossed by internal shocks are
generally assumed to be highly collimated.
\begin{figure}
\centering
 \includegraphics[width=5in,angle=0]{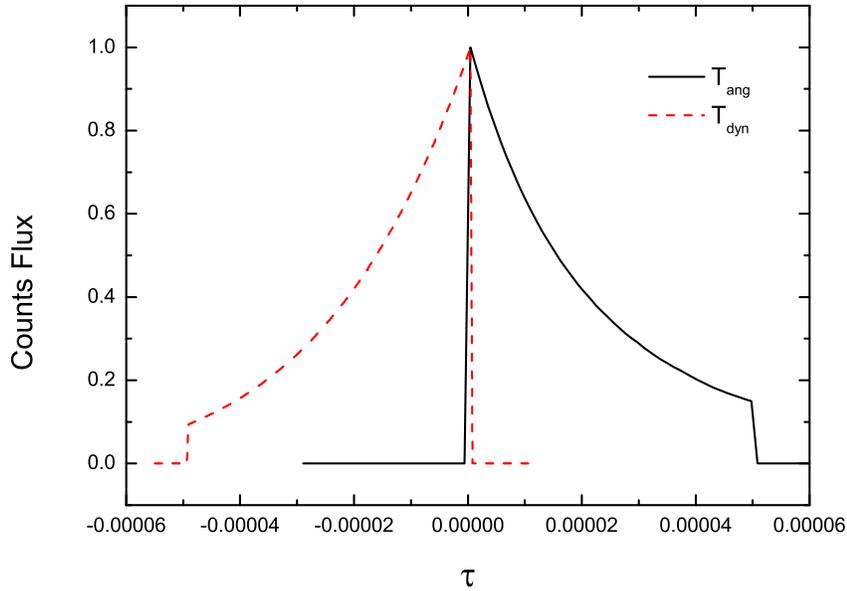}
    \caption{Contrasts between normalized and translated horizontally temporal profiles on timescales
    associated with the curvature effect, $T_{ang}$ with $\Gamma=100$, $R_c=10^{13} cm$,
    $\theta=1/\Gamma$ and $\Delta\tau_{\theta}=0.001$ and the shell-crossing time, $T_{dyn}$
    with $\Gamma=100$, $R_c=10^{13} cm$, $\theta=1/(10^5\Gamma)$ and $\Delta\tau_{\theta}=1$.
    Symbols are denoted by solid line for angular spreading time and dashed line for dynamic time in this plot.}
  \label{fig4}
 \end{figure}

As is shown in figure 4, when the width of the rising co-moving
pulse, $\Delta\tau_{\theta}$, becomes larger than 1 or the thickness
of the shell is wide enough, the dynamic timescale would go beyond
the the angular one. The opposite is the angular timescale could be
the leading contribution to observations. If only the two timescales
are comparable in this situation the observed pulses will be
expected to be more symmetric. Or else, the pulses will show the
characteristics of either FRED or ERFD shapes.

\section{Independence of pulse shape on parameters}

The above-mentioned timescales had been proved to be dependent on
the radius, $R$, of the shell$^{32}$. However, what we want to know
now is how the pulses' shape vary with the radius once the lorentz
factors, the width as well as the ratio of co-moving pulses are
definite for distinct radii. These analytic pulses are displayed in
figure 5,
\begin{figure}
\centering
 \includegraphics[width=5in,angle=0]{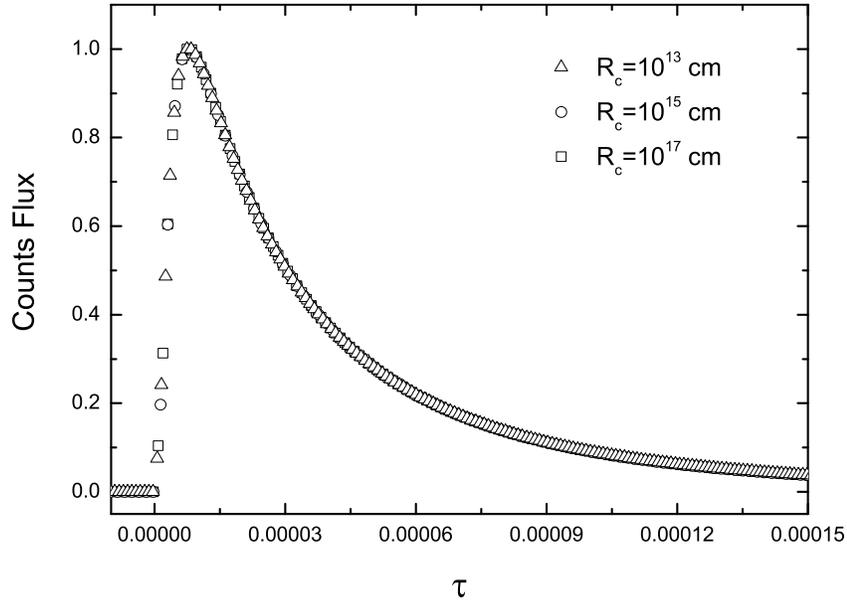}
    \caption{Profiles of the normalized analytic pulses in the observer frame, where I take
    $\Gamma=100$, $\theta=\pi/2$, $\Delta \tau_{\theta}=0.1$ and $\xi=1$.
    Curves for distinct radii are distinguished by triangles, circles and squares respectively.
    Symbols are marked in this plot.}
  \label{fig5}
 \end{figure}
in which we surprisingly see the pulses coming from different radii
are undistinguishable in shape or asymmetry. The consistency
manifests the curves are independent of radii on this occasion,
which in turn shows the parameters, $\Gamma$, $\Delta \tau_{\theta}$
and $\xi$, or part of them, should evolve with radius rather than
keep constant.

It had been known that the bulk Lorentz factor increases linearly
with radius so long as the fireball is not baryon loaded and not
complicated by non-spherical expansion$^{7}$ until $\Gamma\sim1000$
$^{35, 40}$, and then follows a $\Gamma\propto t^{-3/8}$ law
$^{21}$. Unfortunately, the initial value of the shell width
$\Delta^{'}$ (or $t_{dyn}^{'}$) together with its evolution with
radius hasn't been understood yet. I assume the co-moving width
decreases with the increasing of radius, namely,
$\Delta\tau_{\theta}\propto1/R$, because of the violent interaction
of the shells with circum-burst medium. For example, the lorentz
factors are taken as $\Gamma=100, 1000, 10$ whose corresponding
values of other parameters are $\Delta\tau_\theta=10, 1, 0.01$ and
$R_c=10^{12} cm, 10^{13} cm, 10^{15} cm$ respectively. The expected
pulses in observer frame are presented in figure 6, where they are
identified by different symbols. Figure 6 seems to show the observed
pulses arising from larger radius regions would be more asymmetric
and be more close to the property of FRED. This might happen when
the angular spreading time exceeds the dynamic one on condition that
the two factors are considered.
 \begin{figure}
\centering
 \includegraphics[width=5in,angle=0]{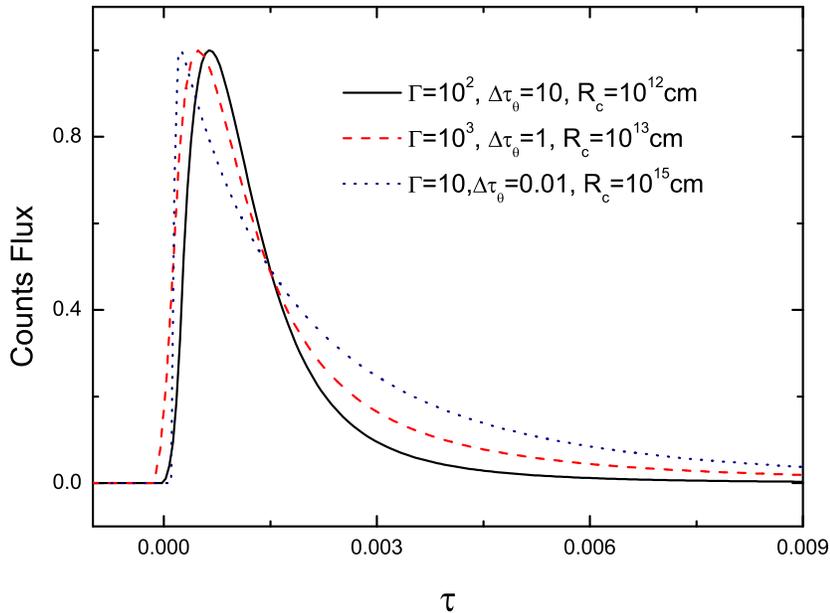}
    \caption{Plot of counts flux vs. $\tau$ for the normalized and re-scaled curves with the different sets of parameters
     of which their symbols have been shown in this plot. Besides, $\theta=\pi/2$ and $\xi=1$ are also designated.}
  \label{fig6}
 \end{figure}

\section{Constraints on the intrinsic times}

Once the cosmological effect is taken into account, the pulse
duration should then be determined by
\begin{equation}
T_{dur}\sim(1+z)T\sim(1+z)(1+\Delta\tau_{\theta})R_c/2\Gamma^2c
\end{equation}
where $z$ is the cosmological redshift for one given source.
Previous investigations$^{3, 42}$ had shown that the spectral lag is
a direct consequence of spectral evolution. During the time interval
of burst, in principal, the full photons emitted from distinct
regions would contribute to the time lag. In the frame of internal
shocks, the temporal and spectral features of pulses are probably
governed by the hydrodynamics process instead of the curvature
effect of the fireball surface$^{3}$. In that case, the lag depends
mainly on the dynamic timescale. On the other hand, either cooling
time$^{30}$ or angular spreading time$^{31}$ can also separately
results in spectral lag. Although Ryde$^{43}$ has suggested the lag
is mainly caused by the pulse decay-time, we here consider the whole
contribution of the above three timescales to time lag lest
additional errors could be produced. From eq.(10), we can easily get
\begin{equation}
\frac{T_{dur, s}}{T_{dur, l}}
\sim\frac{(1+z_s)(1+\Delta\tau_{\theta, s})R_{c,
s}\Gamma_{l}^2}{(1+z_l)(1+\Delta\tau_{\theta, l})R_{c,
l}\Gamma_{s}^2}
\end{equation}
where the subscripts $l$ and $s$ respectively denote long and short
bursts.

Norris \& Bonnell (2006) have shown the median lag is about 48\, ms
for long bursts and less than 1\, ms for short ones. Assuming the
leading contribution to pulse duration is the angular spreading
time, $T_{\mathrm{dur}} \sim(1+z)R_{\mathrm{c}}/2\Gamma^{2}c$, i.e.
lag, we can thus rewrite eq. (11) as
\begin{equation}
\label{eq-12}
\frac{(1+z_{\mathrm{s}})\Gamma_{\mathrm{l}}^{2}}{(1+z_{\mathrm{l}})\Gamma_{\mathrm{s}}^{2}}\frac{(1+\Delta\tau_{\theta,
\mathrm{s}})R_{\mathrm{c, s}}}{(1+\Delta\tau_{\theta,
\mathrm{l}})R_{\mathrm{c, l}}}\leq\frac{1}{48}
\end{equation}
where the median redshifts are respectively appointed to be
$z_{\mathrm{s}}\simeq0.5,\,z_{\mathrm{l}}\simeq2.5$ for short and
long bursts (Norris \& Bonnell 2006). After submitting the values of
redshift to eq.(12), one can derive
\begin{equation}
\label{intrinsic-time} \frac{(1+\Delta\tau_{\theta,
\mathrm{l}})R_{\mathrm{c, l}}}{(1+\Delta\tau_{\theta,
\mathrm{s}})R_{\mathrm{c,
s}}}\geq20\frac{\Gamma_{\mathrm{l}}^{2}}{\Gamma_{\mathrm{s}}^{2}}
\end{equation}
from which we can estimate the lower limit of lorentz factors for
short bursts once the accurate limits of $\Gamma_{\mathrm{l}}$ can
be well known. Some authors had determined the relatively exact
value of the Lorentz factor for long GRBs, by using the reverse
shock information, whose typical values range from 100 to 1000
$^{e.g. 51, 52}$. However, the information of the local pulse width,
$\Delta\tau_{\theta}$, and the radius, $R_{\mathrm{c}}$, hasn't been
achieved until now. Therefore, before this estimation, the width and
the radius need to be reasonably assigned or assumed in advance. In
particular, eq. (13) shows that the intrinsic time,
$t^{\mathrm{'}}=\Delta\tau_{\theta}R_{\mathrm{c}}/c$, is decided by
the ratios of lorentz factors and radii of shells between short and
long bursts.

\section{Conclusion and discussion}

By studying the influences of different timescales on the shape of
pulses, we conclude that the profiles of observed pulses arising
from angular spreading times would be long-tailed and called
standard decay form without rising part, and those only resulting
from dynamic timescales would follow a rising form. Which one is
more dominant than another is determined by the value of
$\Delta\tau_{\theta}$, i.e. $\Delta\tau_{\theta}>1$ for dynamic
dominant case, and vice versa. The pure radiative cooling times
would lead to smooth and FRED-like temporal profiles without too
long tails. Additionally, we find the intrinsic emission time,
$t^{\mathrm{'}}$, is constrained by the ratios of lorentz factors
and radii of the shells between short and long bursts.

Spada et al.$^{32}$ found by simulating internal shocks that the
angular spreading and the dynamic times are comparable when a shell
broadens linearly in some special regions. When it happens, the
thickness $\Delta^{'}$ of the shell could be estimated by
$\Delta^{'}\sim R_c v'_{sh}/(\Gamma c)$, which offer us a clue to
understand that $T_{dyn}$ will go beyond $T_{ang}$ as $\Delta^{'}\gg
R_cv'_{sh}/(\Gamma c)$. In this case, this conclusion just meet
those previous viewpoints that the effect of not the geometry but
the hydrodynamics governs the temporal and spectral characteristics
of GRB pulses$^{3}$. On the contrary, the component of duration from
$T_{ang}$ would always be larger than from $T_{dyn}$ no matter how
large the radii of fireball are.

On the other hand, for larger radii the radiative cooling time will
be the dominant contribution to the pulse duration$^{32}$, thus the
shape of observed pulses should reflect the properties dominated by
radiation during the whole duration. Moreover, the lorentz factors,
thickness of the merged shell, the frequency of emitted photons and
the energy equipartition factor $\varepsilon_{B}$ of the magnetic
field will reduce to smaller values with the increasing of radii. On
this occasion, there is a more strong contribution from $T_{syn}$ to
observed pulses in contrast with either $T_{ang}$ or $T_{syn}$. Even
if the outflows ejected from central engine are extremely calibrated
in the early phase, the jet will spread sideways quickly$^{20}$ so
that its geometry effects on observations are made to be more
considerable. Therefore, the observed pulses are caused to exhibit
more asymmetric and the trend of FRED are then strengthened.

We have known the leading model of central engine for long GRBs,
perhaps including short ones, could be the collapsar model$^{36,
38}$. Besides, other merger models of two compact objects have also
been proposed$^{44, 45, 46, 47}$. However, the analysis in this work
is independent of the detailed progenitor models. The lorentz
factors had been estimated as an order of 100-1000 for long
bursts$^{51, 52}$, while its lower limit is about 500 for short ones
(Norris and Bonnell 2006). Meanwhile, we have reached a conclusion
in our recent works$^{48, 49}$ that the lorentz factors is
proportional to $\Gamma^{-\omega}$ with $\omega>2$ for short bursts
and $\omega<2$ for long bursts. In terms of the properties of tiny
spectral lags in short bursts as well as their symmetric pulse
characteristic, I thus infer that, although short and long GRB
pulses could be interpreted with the same emission mechanism$^{12}$
in terms of the possible synchrotron radiation, the long and
asymmetric FRED pulses could be produced by external shock in larger
radii$^{14}$, while the short and symmetric pulses might be formed
by internal shock in smaller emission distance from sources. We
expect it to be verified in the future by the SWIFT satellite in
view of its capability of precise and prompt localization.



\end{document}